\documentclass[prd, onecolumn, nofootinbib]{revtex4}

\usepackage{graphics}
\usepackage{amsmath}

\def\dblone{\hbox{$1\hskip -1.2pt\vrule depth 0pt height 1.6ex width 0.7pt
                  \vrule depth 0pt height 0.3pt width 0.12em$}}
\def\eq#1{Eq.~(\ref{#1})}

\begin{document}

\title{Proof Of The Invalidity Of The Boltzmann Property In The FMO Many-Body Neutrino Model}

\author{James Quach} \email{jamesq@unimelb.edu.au}
\affiliation{School of Physics, The University of Melbourne, Victoria 3010, Australia}


\begin{abstract}
There has been recent debate over the use of the Boltzmann property in the kinetic equations describing dense neutrino systems such as early Universe and Supernova core. A technique developed by Bell, Rawlinson, and Sawyer utilises the flavour evolution timescales of the neutrino systems to test the validity of this assumption. The Friedland-McKellar-Okuniewicz (FMO) many-body neutrino model was developed to conduct this test. It was concluded by its authors, using the Bell-Rawlinson-Sawyer timescale test, that the model lent support to the Boltzmann property assumption. We developed kinetic equations for the FMO model. By direct analysis of the kinetic equations we find, in stark contrast to Friedland et al., that in fact the Boltzmann property assumption does breakdown in the FMO model. We have shown that the Bell-Rawlinson-Sawyer timescale technique can only be used to invalidate the Boltzmann property but not validate it.
\end{abstract}


\maketitle


\section{Introduction}
\label{Introduction}

This paper is based on the author's 2006 University of Melbourne BSc honour's report supervised by Bruce H.J. McKellar.\\

Single-body kinetic equations which encompass neutrino oscillation as well as neutrino-neutrino scattering have been found by Pantaleone \cite{pantaleone92},\cite{pantaleone92a}, Sigl and Raffelt \cite{sigl93}, McKellar and Thomson \cite{mcKellar94} to describe neutrino dynamics in early Universe and Supernovae. Underpinning these kinetic equations is the typical assumption, originally put forth by Boltzmann for classical dilute gases, that the neutrino density operator can be expressed as a tensor product of the individual one-body density operators of the system i.e.

\begin{equation}
	\rho=\prod_{n=1}^N\otimes\rho_n
\label{eq:oneApprox}
\end{equation}

The validity of this assumption is not quantum mechanically obvious as interactions between neutrinos may produce entangled states which are not products of individual wavefunctions. 

The effects of entanglement have up until now been looked from two approaches. The first by Friedland and Lunardini \cite{friedland03}, constructed the flavour evolution of a many-body neutrino system as an effect of the interference of many elementary neutrino-neutrino scattering events. With this construction, they argued that in the limit of an infinite number of neutrinos the system can be factorised into one-body states.

The alternate approach, introduced by Bell et al. \cite{bell03}, studies the rate at which the many-body neutrino system reaches statistical equilibrium. Essentially the argument here is that the one-body\footnote{By \emph{one-body} we mean \emph{factorisation into one-body states}. This is an abbreviation used in the literature, which we will also use.} description predicts equilibration to be reached in time scales of an incoherent\footnote{Here, coherent and incoherent refers to the type of scattering} nature. Bell found equilibration to be reached in time scales of a coherent nature, thus citing evidence of the breakdown of the one-body description. Subsequently, Friedland and Lunardini \cite{friedland03a} found for systems with two types of neutrinos in equal number, equilibration time scales to be of an incoherent nature, providing support for the one-body description. Later, Friedland, McKellar, Okuniewicz \cite{friedland06}, generalised this model, and concluded equilibration time scales to be the incoherent type. This latter model is what we call the Friedland-McKellar-Okuniewicz many body neutrino model or FMO model for short. Because equilibration times scales were of an incoherent nature Friedland et. al. found, at least in the FMO model, no evidence for the failure of the one-body description.

In this paper we enquire about the vailidity of the one-body description from a different angle. At the heart of the one-body description is the assumption that correlations between particles are insignificant. The question that naturally arises then is exactly what role does correlation affects have and are they indeed insiginificant. In an attempt to answer this question we develop an exact kinetic equation of the Bogolubov-Born-Green-Kirkwood-Yvon (BBGKY) type. This will be used to measure the correctness of the FMO calculations. If they are correct, their results should exactly satisfy the exact kinetic equation. Then incorporating the one-body approximation, a Boltzmann-like kinetic equation will be developed. The degree to which this approximate equation is satisfied in the FMO model will then be investigated. The intention is that it will yield insights into the role of the correlation matrix and the validity of the one-body approximation in the FMO model. 


\section{Friedland-McKellar-Okuniewicz Many Body Neutrino Model}
\label{Setup}

The FMO Model is a `toy' model that was used by its authors to analyse the flavour evolution of a system of interacting neutrinos. So that the system could be exactly solved the authors made the simplifications that there are only two neutrino types and no spatial dependence. Although obviously this is not a representation of reality, it allowed the authors to exactly investigate the dynamics of flavour conversion. In this section we derive the hamiltonian and the one and two-body density matrices of the FMO many-body neutrino model. 

\subsection{Hamiltonian Of The FMO Model}
\label{sec:Hamiltonian_FMO}

The hamiltonian of the FMO model presented in this section is a reproduction of \cite{friedland06} albeit liberties with notation.

In the FMO model, neutrino flavour is represented as spin. The interaction between neutrinos is described by the low-energy neutral current Hamiltonian
\begin{equation}
	H_{NC}=\frac{G_F}{\sqrt{2}}\left(\sum_\alpha{j^\mu_\alpha}\right)\left(\sum_\beta{j_{\mu\beta}}\right)
\label{eq:ncInteraction}
\end{equation}

\begin{tabbing}
where \= $j^\mu_\alpha\equiv\bar{\nu_\alpha}\gamma^\mu\nu_\alpha$\\
	\>$\alpha=1..k$ is a flavour index\\
	\>$G_F$ is the Fermi constant.
\end{tabbing}

Because this Hamiltonian is invariant under the rotation of the SU(2) flavour group the flavour space is isomorphic to spin interactions.

Consider a two neutrino interaction. Since neutral current interactions conserve flavour, in the Hamiltonian of (\ref{eq:ncInteraction}) the flavour space wavefunction of a given outgoing neutrino is equal to that of one of the incoming neutrinos. Denoting the flavour wavefunctions as $\Psi_i$ and $\Phi_i$, the only possible combinations are

\begin{align}
	\Psi_i^*\delta_{ij}\Psi_j\Phi^*_k\delta_{kl}\Phi_l \label{eq:interactionCombo2}\\	
	\Psi_i^*\delta_{il}\Phi_l\Phi^*_k\delta_{kj}\Psi_j \label{eq:interactionCombo1}
\end{align}

where we have suppressed the gamma matrices for clarity.

Using the fact that $\sum^2_{i=1}|\Psi_i|^2=\sum^2_{i=1}|\Phi_i|^2=1$ and the relation $2\delta_{il}\delta_{jk}=\delta_{ij}\delta_{kl}+\vec{\sigma_{ij}}\cdot\vec{\sigma_{kl}}$, we can write the sum of (\ref{eq:interactionCombo1}) and (\ref{eq:interactionCombo2}) as

\begin{equation}
	\frac{3}{2}+\frac{\vec{\sigma_{1}}\cdot\vec{\sigma_{2}}}{2}
\label{eq:spinInteraction}
\end{equation}

Thus the pair-wise interaction Hamiltonian can be written as,

\begin{equation}
	\begin{split}
		H_{ij}&=g(\frac{3}{2}+\frac{\vec{\sigma_{i}}\cdot\vec{\sigma_{j}}}{2})\\
			&=g(\frac{\vec{\sigma}_i}{2}+\frac{\vec{\sigma}_j}{2})^2\\
	\end{split}
\end{equation}

where $g$ is a coupling constant.

The interaction Hamiltonian for the whole system of $N$ neutrinos is then,

\begin{equation}
	\begin{split}
	H&=g\sum_{i=1}^{N-1}\sum_{j=i+1}^{N}\left(\frac{\vec{\sigma_i}}{2}+\frac{\vec{\sigma_j}}{2}\right)^2\\
	&=g\left(\sum_{i=1}^{N-1}\sum_{j=i+1}^{N}\frac{\vec{\sigma_i}\cdot\vec{\sigma_j}}{2}+\frac{3}{4}N(N-1)\right)\\
	&=g\left(J^2+\frac{3}{4}N(N-2)\right)
	\end{split}
\label{eq:interactionHamiltonian}
\end{equation}

where we have used the fact that 

\begin{equation}
	\hat{J}^2 = \left( \sum_{i=1}^{N}{\frac{\vec{\sigma_i}}{2}} \right)^2 = \sum_{i=1}^{N-1} \sum_{j=i+1}^{N} \frac{\vec{\sigma_i}\cdot\vec{\sigma_j}}{2} + \frac{3}{4}N
\end{equation}

\subsection{The One And Two Body Density Matrices Of The FMO Model}
\label{sec:One_Two_Density_Matrices_FMO}

The original derivation of the one-body density matrix in the FMO model was conducted in \cite{friedland06}. We reproduce the result here and extend it to the two-body case. 

\subsubsection{Derivation Of The One Body Density Matrix}
\label{sec:DerivationOneBody}

We begin by considering the Hamiltonian of the system (\ref{eq:interactionHamiltonian}). Its eigenvalues are

\begin{equation}
	H\Psi_{J,M}=g\Bigl( J(J+1) + \frac{3}{4}N(N-2) \Bigr) \Psi_{J,M}
\end{equation}

where $\psi_{J,M}$ is an eigenstate of $H$. The subscripts remind us that the eigenstate is in the total angular momentum basis, $J$, and total angular momentum projection $M$. $N$ is the total number of particles in the system.

Let our system begin in the state

\begin{equation}
	|j_{U} m_{U}>\otimes|j_{D} m_{D}>=|j_{U} j_{D};m_{U} m_{D}>
\label{eq:intialState}
\end{equation}

where $|j_{U} m_{U}>$ is the total state of all spin-up particles
and $|j_{D} m_{D}>$ is the total state of all spin-down particles.

Now the interaction Hamiltonian (\ref{eq:interactionHamiltonian}) will evolve this state in time. To do this conveniently, we wish to write our state in the total angular momentum basis,

\begin{equation}
	|j_{U} j_{D};m_{U} m_{D}>=\sum_{J}<j_{U} j_{D};J,M|m_{U} m_{D}> |j_{U} j_{D};J M>
\end{equation}

Now we can easily use $H$ to evolve this system in time,

\begin{equation}
	|\Psi_{J,M}(t)>=\sum_{J}e^{-itE(J)}<J,M|m_{U} m_{D};j_{U} j_{D}> |j_{U} j_{D};J M>
\label{eq:totalBasisState}
\end{equation}

To get the one body density matrix, which we denote as $\rho_{1}$, we need to single out this particular particle. To do this we decouple this one particle from the rest,

\begin{equation}
	|j_{U} j_{D};J M>=|(j_{1} k_{U})j_{U},j_{D};J M>
\end{equation}

where $j_{1}$ is the total angular momentum of the one particle,
$k_{U}$ is the total angular momentum of the rest of the spin-up particles.

Note that we have necessarily specified that the particle we are singling out is initially spin-up. An analogous derivation can be done for a particle that is initially spin-down.

Now using known angular momentum coupling formula \eq{eq:singleDecoupling} we write,
\begin{equation}
	\begin{split}
	|j_{U} j_{D};J M>&=\sum_{j,m_{1},m}(-)^{J+j_{U}+j_{D}}[(2j_{U}+1)(2j+1)]^{\frac{1}{2}}\\
		&\quad\times<j_{1} j;m_{1} m|J M>\left\{
				\begin{array}{ccc}
					j_1 & j_U & j_U \\
					j_D & J & j
				\end{array}\right\}\\
		&\quad\times|j_{1} j;m_{1} m>
	\end{split}
\label{eq:singleDecoupledState}	
\end{equation}

So putting \eq{eq:singleDecoupledState} into \eq{eq:totalBasisState} we have

\begin{equation}
	\begin{split}
	|\Psi_{J,M}(t)>&=\sum_{J,j,m_{1},m}(-)^{J+j_U+J_D}e^{-itE(J)}\\
		&\quad\times[(2j_U+1)(2j+1)]^{\frac{1}{2}}\\
		&\quad\times<j_U j_{D};J M|m_U m_D> <j_1 j; J M|m_1 m>\\
		&\quad\times\left\{
				\begin{array}{ccc}
					j_1 & j_U & j_N \\
					j_D & J & j
				\end{array}\right\}|j_{1} j;m_{1} m>
	\end{split}
\end{equation}

The density matrix of the total system is then,

\begin{equation}
	\begin{split}
			\rho(t)&=|\Psi_{J,M}(t)><\Psi_{J,M}(t)|\\
			&= \sum_{\substack{J,J',j,j',j_{1},j'_{1}\\m,m',m_{1},m'_{1}}}\exp\{-itg[J(J+1)-J'(J'+1)]\}\\
			&\quad\times(2j_n+1)\sqrt{(2j+1)(2j'+1)}\\
			&\quad\times<j_U j_D; J M | m_U m_D> <j_U j_D; J' M | m_U m_D>\\
			&\quad\times\left\{
				\begin{array}{ccc}
					1/2 & j_U-1/2 & j_U\\
					j_D & J & j
				\end{array}
			\right\}
			\left\{
				\begin{array}{ccc}
					1/2 & j_U-1/2 & j_U \\
					j_D & J' & j'
				\end{array}
			\right\}\\
			&\quad\times|j_1j;m_1m><j_1'j';m_1'm_1|
	\end{split}
\end{equation}

But we want the one-body density matrix. To do this, we take the trace over the rest of the system.

\begin{equation}
	\begin{split}
		\rho_1(t)&=Tr_{|jm>}\rho(t)
	\end{split}
\end{equation}

Thus,

\begin{equation}
	\begin{split}	\rho_1^{(\uparrow)}(t)&=\sum_{\substack{J,J',j,j_{1},j'_{1}\\m,m_{1},m'_{1}}}\exp\{-itg[J(J+1)-J'(J'+1)]\}\\
			&\quad\times(2j_n+1)(2j+1)\\
			&\quad\times<j_U j_D; J M | m_U m_D> <j_U j_D; J' M | m_U m_D>\\
			&\quad\times<j_1 j; J M | m_1 m> <j_1 j; J' M | m'_1 m>\\
			&\quad\times\left\{
				\begin{array}{ccc}
					1/2 & j_U-1/2 & j_N \\
					j_D & J & j
				\end{array}
			\right\}
			\left\{
				\begin{array}{ccc}
					1/2 & j_U-1/2 & j_N \\
					j_D & J' & j
				\end{array}
			\right\}\\
			&\quad\times|j_1 m_1><j_1 m'_1|
	\end{split}
\label{eq:oneBodyDensityMatrixUp}
\end{equation}

The superscript $(\uparrow)$ in \eq{eq:oneBodyDensityMatrixUp} indicates that the particle denoted by the subscript, in this case particle 1, was initially in the spin-up state. For a particle that is initially down, one need only replace the $U$ with $D$ and vice versa, i.e.,

\begin{equation}
\label{eq:oneBodyDensityMatrixDown}
\begin{split}	\rho_1^{(\downarrow)}(t)&=\sum_{\substack{J,J',j,j_{1},j'_{1}\\m,m_{1},m'_{1}}}\exp\{-itg[J(J+1)-J'(J'+1)]\}\\
		&\quad\times(2j_n+1)(2j+1)\\
		&\quad\times<j_D j_U; J M | m_D m_U> <j_D j_U; J' M | m_D m_U>\\
		&\quad\times<j_1 j; J M | m_1 m> <j_1 j; J' M | m'_1 m>\\
		&\quad\times\left\{
			\begin{array}{ccc}
				1/2 & j_D-1/2 & j_N \\
				j_U & J & j
			\end{array}
		\right\}
		\left\{
			\begin{array}{ccc}
				1/2 & j_D-1/2 & j_N \\
				j_U & J' & j
			\end{array}
		\right\}\\
		&\quad\times|j_1 m_1><j_1 m'_1|
\end{split}
\end{equation}

\subsubsection{Derivation Of The Two Body Density Matrix}
In this section we derive the two-body density matrix. If the two particles are initially spin aligned, the derivation follows that of \S\ref{sec:DerivationOneBody}, with the difference being that one must decouple two particles instead of one. The final result is that one need only make the following changes to \eq{eq:oneBodyDensityMatrixUp},
\begin{itemize}
\item replace $j_1$ with $j_{12}$ where $j_{12}$ is the summation of the angular momentum of the two particles i.e. $j_{12}=j_1+j_2$
\item replace $m_1$ with $m_{12}$ where $m_{12}=-j_{12},-j_{12}+1,...,j_{12}-1,j_{12}$
\end{itemize}

\begin{equation}
\begin{split}
		\rho_{12}^{(\uparrow\uparrow)}(t)&=\sum_{\substack{J,J',j,j_{1},j'_{1}\\m,m_{1},m'_{1}}}\exp\{-itg[J(J+1)-J'(J'+1)]\}\\
		&\quad\times(2j_n+1)(2j+1)\\
		&\quad\times<j_U j_D; J M | m_U m_D> <j_U j_D; J' M | m_U m_D>\\
		&\quad\times<j_{12} j; J M | m_{12} m> <j_{12} j; J' M | m'_1 m>\\
		&\quad\times\left\{
			\begin{array}{ccc}
				1/2 & j_U-1/2 & j_N \\
				j_D & J & j
			\end{array}
		\right\}
		\left\{
			\begin{array}{ccc}
				1/2 & j_U-1/2 & j_N \\
				j_D & J' & j
			\end{array}
		\right\}\\
		&\quad\times|j_{12} m_{12}><j_{12} m'_1|
\end{split}
\label{eq:twoBodyDensityMatrixUp_1}
\end{equation}

\begin{equation}
	\begin{split}	
	\rho_{12}^{(\downarrow\downarrow)}(t) &= \sum_{\substack{J,J',j,j_{1},j'_{1}\\m,m_{1},m'_{1}}}\exp\{-itg[J(J+1)-J'(J'+1)]\}\\
			&\quad\times(2j_n+1)(2j+1)\\
			&\quad\times<j_D j_U; J M | m_D m_U> <j_D j_U; J' M | m_D m_U>\\
			&\quad\times<j_{12} j; J M | m_{12} m> <j_{12} j; J' M | m'_1 m>\\
			&\quad\times\left\{
				\begin{array}{ccc}
					1/2 & j_D-1/2 & j_N \\
					j_U & J & j
				\end{array}
			\right\}
			\left\{
				\begin{array}{ccc}
					1/2 & j_D-1/2 & j_N \\
					j_U & J' & j
				\end{array}
			\right\}\\
			&\quad\times|j_{12} m_{12}><j_{12} m'_1|
	\end{split}
\label{eq:twoBodyDensityMatrixDown}
\end{equation}

The essence of the derivation for particles that are initially in opposite spins is to decouple one particle from the spin up system, and one from the spin-down system. To do this, we use the following three relations,

\begin{equation}
	|j_U j_D;J M> = \sum_{j_{12},j}<j_U j_D;J M|j_{12} j; J M> |j_{12} j;J M>
\label{eq:decoupleTwoParticles}
\end{equation}

\begin{equation}
	|j_{12} j, J M> = \sum_{m_{12},m}<j_{12} j; m_{12} m|j_{12} j; J M> |j_{12} j; m_{12} m>
\label{eq:changeTwoParticleBasis}
\end{equation}

And with the help of \eq{eq:doubleDecoupling},

\begin{equation}
	\begin{split}
	&<(j_1 k_U)j_U,(j_2,k_D)j_D;J M|(j_1 j_3)j_{12},(k_U k_D);J M> =\\
		&\quad[(2j_U+1)(2j_D+1)(2j_{12}+1)(2j+1)]^{\frac{1}{2}}
		\left\{
				\begin{array}{ccc}
					j_1 & k_U & j_U \\
					j_2 & k_D & j_D\\
					j_{12} & j & J\\
				\end{array}
			\right\}
	\end{split}
\label{eq:decoupleEachParticle}
\end{equation}	

If we put the above three relations into \eq{eq:totalBasisState}, and  recall that $\rho(t)=|\Psi_{J,M}(t)><\Psi_{J,M}(t)|$, we now have a density matrix in a form which is trivial to extract the two-body density matrix by tracing over the rest of the system. The final result is,

\begin{equation}
	\begin{split}
		\rho_{12}^{(\uparrow\downarrow)}(t)&=\sum_{\substack{J,J',j,j_{12},j'_{12}\\m,m_{12},m'_{12}}}\exp\{-itg[J(J+1)-J'(J'+1)]\}\\
		&\quad\times(2j_U+1)(2j_D+1)(2j+1)[(2j_{12}+1)(2j'_{12}+1)]^{\frac{1}{2}}\\
		&\quad\times<j_U j_D; J M | m_U m_D> <j_U j_D; J' M | m_U m_D>\\
		&\quad\times<j_{12} j; m_{12} m |J M> <j'_{12} j; m'_{12} m |J' M>\\
		&\quad\times\left\{
			\begin{array}{ccc}
				j_1 & k_U & j_U \\
				j_2 & k_D & j_D\\
				j_{12} & j & J\\
			\end{array}
		\right\}
		\left\{
			\begin{array}{ccc}
				j_1 & k_U & j_U \\
				j_2 & k_D & j_D\\
				j_{12} & j & J'\\
			\end{array}
		\right\}\\
	&\quad\times|j_{12} m_{12}><j'_{12} m'_{12}|\\
	\end{split}
\label{eq:twoBodyDensityMatrix}
\end{equation}

\subsection{Summary Of Parameter Definitions And Density Matrix Notation}
\label{sec:summaryDef}

For convenience we summarise here the parameters used to describe the FMO model.
\begin{itemize}
\item $J$ is the total angular momentum of the system
\item $M$ is the the projection of $J$
\item $j$ is the total angular momentum of the rest of the system e.g. in the one-body density matrix it is the total angular momentum of the system less particle 1; for the two-body density matrix is is the total angular momentum of the system less particle 1 and particle 2.
\item $m$ is the angular momentum projection of $j$
\item $j_U$ is the total angular momentum of all the spin-up particles in the system
\item $m_U$ is the the projection of $j_U$
\item $j_D$ is the total angular momentum of all the spin-up particles in the system
\item $m_D$ is the the projection of $j_D$
\item $k_U$ is the total angular momentum of the rest of the spin-up system e.g. in the one-body density matrix where particle 1 is initially spin-up, it is the total angular momentum of the system less particle 1; for the two-body density matrix of where particle 1 and 2 are initially spin-up, it is the total angular momentum of the spin-up system less particle 1 and particle 2.
\item $k_D$ is the total angular momentum of the rest of the spin-down system e.g. in the one-body density matrix where particle 1 is initially spin-down, it is the total angular momentum of the system less particle 1; for the two-body density matrix of where particle 1 and 2 are initially spin-down, it is the total angular momentum of the spin-down system less particle 1 and particle 2.
\item $N$ is the total number of particles in the system
\item $g$ is a coupling constant
\end{itemize}


\section{Kinetic Equation In FMO Model}
\label{sec:ChKineticEq}

Classically the time evolution of the phase space distribution is described by the Liouville equation,

\begin{equation}
	\frac{\partial \rho}{\partial t}=\{H, \rho\}
\end{equation}

where H is the Hamiltonian of the system, and curly brackets are Poisson brackets.

The quantum mechanical analogue is the time evolution of the density matrix, which is described by the Von Neumann equation,

\begin{equation}
	i \frac{\partial \rho}{\partial t}=[H,\rho]
\end{equation}

We can split the Hamiltonian into its single particle and interaction constituents,

\begin{equation}
	H=\sum^N_{i=1}{K_i}+\frac{1}{2}\sum_{i\neq j}V_{ij}
\end{equation}

where $K$ is the single particle Hamiltonian of the $i$th particle and $V_{ij}$ is the interaction Hamiltonian between the $i$th and $j$th particle.

The rate of change of a single particle density matrix is found by tracing over all the other particles,

\begin{equation}
\label{eq:bbgky}
	\begin{split}
		i \frac{\partial \rho_1}{\partial t}&=[K_1,\rho_1]+ \frac{1}{2}\sum_{i\neq j}\mbox{Tr}_{2..N}[V_{ij},\rho_{1..N}]\\
		&=[K_1,\rho_1]+\frac{1}{2}\sum_j \mbox{Tr}_j[V_{1j},\rho_{1j}]+\frac{1}{2}\sum_i \mbox{Tr}_i[V_{i1},\rho_{1i}]\\
		 &\quad+\frac{1}{2}\sum_{i\neq j} \mbox{Tr}_{ij}[V_{ij},\rho_{1ij}]\\
		&=[K_1,\rho_1]+\sum_j \mbox{Tr}_j[V_{1j},\rho_{1j}]+\frac{1}{2}\sum_{i\neq j} \mbox{Tr}_{ij}[V_{ij},\rho_{1ij}]
	\end{split}
\end{equation}

In the last line of \eq{eq:bbgky} we have employed the symmetry of our interaction Hamiltonian.

\eq{eq:bbgky} is in fact the quantum analogue of the first of the Bogolubov-Born-Green-Kirkwood-Yvon (BBGKY) equations\cite{huang87}.

In the FMO model there is no single particle Hamiltonian, so we will omit it.

For the interaction Hamiltonian of our model it can readily be shown that the last term goes to zero. Here is the proof.

\begin{widetext}
\begin{equation}
	\begin{split}
		\mbox{Tr}_{ij}[V_{ij},\rho_{1ij}]&=\mbox{Tr}_{ij}(<a_1 a_i a_j|V_{ij} \rho_{1ij}|a'_1 a'_i a'_j>-<a_1 a_i a_j|\rho_{1ij} V_{ij}|a'_1 a'_i a'_j>)\\
		&=\sum_{a_i a_j}{<a_1 a_i a_j|V_{ij} \rho_{1ij}|a'_1 a'_i a'_j>-<a_1 a_i a_j|\rho_{1ij} V_{ij}|a'_1 a'_i a'_j>}\\
		&=\sum_{\substack{a_i,a_j\\b_1,b_i,b_j}}<a_1 a_i a_j|V_{ij}|b_1 b_i b_j><b_1 b_i b_j| \rho_{1ij}|a'_1 a'_i a'_j>\\
		&\quad-<a_1 a_i a_j|\rho_{1ij}|b_1 b_i b_j><b_1 b_i b_j|V_{ij}|a'_1 a'_i a'_j>
	\end{split}
\end{equation}

where summation over primed equivalents is implied.
\end{widetext}

With the appropriate choice of an orthonormal basis one can see that $a_1=b_1=a'_1$, otherwise the term is zero. So,

\begin{equation}
	\begin{split}
		Tr_{ij}[V_{ij},\rho_{1ij}]&=\sum_{\substack{a_i,a_j\\b_i,b_j}}<a_1 a_i a_j|V_{ij}|a_1 b_i b_j><a_1 b_i b_j| \rho_{1ij}|a_1 a'_i a'_j>\\
		&\quad-<a_1 a_i a_j|\rho_{1ij}|a_1 b_i b_j><a_1 b_i b_j|V_{ij}|a_1 a'_i a'_j>
	\end{split}
\end{equation}

Using the fact that our interaction Hamiltonian is symmetric and that the appropriate basis consists of only two states for each particle, namely spin-up or spin-down, the summation over all possible $a_i,a_j,b_i,b_j$, will yield zero as the negative sign in the second term will ensure that every term will cancel.

Let's now consider the second term in \eq{eq:bbgky} i.e. $\sum_j \mbox{Tr}_j[V_{1j},\rho_{1j}]$. In this term there are two types of two-body density matrices: that where the spin of the two particles are initially aligned and that where the spin of the two particles are initially anti-aligned. Thus, \eq{eq:bbgky} can be simplified to,

\begin{equation}
	i \frac{\partial \rho_1}{\partial t}=N Tr_2[V_{12},\rho_{12}]+M Tr_3[V_{13},\rho_{13}]
\label{eq:simplifiedBBGKY}	
\end{equation}

where we have labelled a particle that is initially anti-aligned to particle 1 as particle 2 and a particle that is initially aligned as particle 3. $N$ is the number of initially anti-aligned particles to particle 1, and $M$ is the number of initially aligned-particles to particle 1. Note that $N$ here is different from $N$ in \S\ref{sec:summaryDef}, which was defined as the total number of particles.

In the next two sections we show that particles that are initially aligned to particle 1 do no contribute to its time evolution, and particles that are initially anti-aligned to particle 1 do.

\subsection{Initially Aligned Particles Do Not Contribute To Kinetic Equation}
\label{Initially_Aligned}

The interaction Hamiltonian in matrix form is,

\begin{equation}
	V_{ij}=\frac{3g}{2}\dblone+\frac{g}{2}
	\begin{pmatrix}
	1&0&0&0\\
	0&-1&2&0\\
	0&2&-1&0\\
	0&0&0&1
	\end{pmatrix}
\label{eq:Vij4}	
\end{equation}

where $\dblone$ is the unit matrix.

The spin basis of \eq{eq:Vij4} reading left to right (and of course top to bottom) is, $|\uparrow\uparrow>$,$|\uparrow\downarrow>$,$|\downarrow\uparrow>$,$|\downarrow\downarrow>$, where the first arrow represents the spin of particle one, and the second the spin of particle two.

Let us now define a general two-body density matrix,
\begin{equation}
\label{eq:rho1n4}
\rho_{1n}\equiv
\begin{pmatrix}
\rho^{11}&\rho^{12}&\rho^{13}&\rho^{14}\\
\rho^{21}&\rho^{22}&\rho^{23}&\rho^{24}\\
\rho^{31}&\rho^{32}&\rho^{33}&\rho^{34}\\
\rho^{41}&\rho^{42}&\rho^{43}&\rho^{44}\\
\end{pmatrix}
\end{equation}
\eq{eq:rho1n4} has the same basis as \eq{eq:Vij4}.
Using \eq{eq:Vij4} and \eq{eq:rho1n4} we evaluate that,
\begin{equation}
\label{eq:alignedTerm}
{Tr}_n[V_{1n},\rho_{1n}]=g
\begin{pmatrix}
	\rho^{32}-\rho^{23}&\rho^{13}-\rho^{12}+\rho^{34}-\rho^{24}\\
	\rho^{21}-\rho^{31}+\rho^{42}-\rho^{43}&\rho^{23}-\rho^{32}
\end{pmatrix}
\end{equation}

The two-body density matrix $\rho_{13}$ of \eq{eq:simplifiedBBGKY} in matrix element form is,
\begin{equation}
	\begin{split}
			&<m_1,m_3|\rho_{13}(t)|m_1',m_3'>\\
	&\quad=\sum_{\substack{J,J',j,m}}(-)^{J-J'}\exp\{-itg[J(J+1)-J'(J'+1)]\}\\
			&\quad\times(2j_U+1)(2j+1)\\
			&\quad\times<j_U j_D; J M | m_U m_D> <j_U j_D; J' M | m_U m_D>\\
			&\quad\times<j_{13} j; J M | m_{13} m> <j_{13} j; J' M | m'_{13} m>\\
			&\quad\times\left\{
				\begin{array}{ccc}
					j_{13} & j_U-j_{13} & j_U \\
					j_D & J & j
				\end{array}
			\right\}
			\left\{
				\begin{array}{ccc}
					j_{13} & j_U-j_{13} & j_U \\
					j_D & J' & j
				\end{array}
			\right\}
	\end{split}
\label{eq:twoBodyDensityMatrixUp_2}
\end{equation}

Now $m_{13}+m=M$. So for element $<\frac{1}{2},\frac{1}{2}|\rho_{13}|\frac{1}{2},-\frac{1}{2}>$,
\begin{equation}
\begin{split}
	&<j_{13} j; m_{13} m |J M> <j_{13} j; m'_{13} m |J' M>\\
	&=<1 j; 1 m |J M> <1 j; 0 m |J' M>\\
	&=0
\end{split}
\end{equation}
since it is not possible for $M$ to satisfy both $m_{13}+m=1+m$ and $m_{13}'+m=0+m$.

Therefore matrix element $\rho^{12}=<\frac{1}{2},\frac{1}{2}|\rho_{13}|\frac{1}{2},-\frac{1}{2}>=0$.
By same reasoning matrix elements, $\rho^{13}=\rho^{34}=\rho^{24}=\rho^{31}=\rho^{21}=\rho^{23}=\rho^{42}=0$. From \eq{eq:alignedTerm} we thus see that the off-diagonal elements of $M \mbox{Tr}_3[V_{13},\rho_{13}]$ are zero.

Let us now make the following definitions,
\begin{align}
\label{eq:a}
a(J,J')&\equiv\sum_{\substack{j,m}}(-)^{J-J'}(2j_U+1)(2j+1)\\
		&\quad\times<j_U j_D; J M | m_U m_D> <j_U j_D; J' M | m_U m_D>\\
		&\quad\times<j_{13} j; J M | m_{13} m> <j_{13} j; J' M | m'_{13} m>\\
		&\quad\times\left\{
			\begin{array}{ccc}
				j_{13} & j_U-j_{13} & j_U \\
				j_D & J & j
			\end{array}
		\right\}
		\left\{
			\begin{array}{ccc}
				j_{13} & j_U-j_{13} & j_U \\
				j_D & J' & j
			\end{array}
		\right\}\\
\label{eq:b}
b(J)&\equiv-itgJ(J+1)
\end{align}
Then $\rho_{13}$ can be written in the following way,
\begin{widetext}
\begin{equation}
\label{eq:twoBodyAligned}
\begin{split}
	\rho_{13}(t)&=\sum_{J,J'}a(J,J')exp\{b(J)-b(J')\}\\
	&=\left(\sum_{J<J'}a(J,J')\exp\{b(J)-b(J')\}+a(J',J)\exp\{b(J')-b(J)\}\right) + \sum_J{a(J,J'=J)}\\
	&=\left(\sum_{J<J'}a(J,J')\{\cos(b(J)-b(J'))-i\sin(b(J)-b(J'))\} + a(J',J)\{\cos(b(J')-b(J))-i\sin(b(J')-b(J))\}\right)\\
	&\quad+ \sum_J{a(J,J'=J)}
\end{split}
\end{equation}
\end{widetext}
Now $a(J,J')$ is symmetric about $J,J'$ i.e. $a(J,J')=a(J',J)$ . Therefore,
\begin{equation}
\label{eq:twoBodyAligned2}
	\rho_{13}(t)=\sum_{J,J'}2a(J,J')cos(b(J')-b(J))+\sum_J{a(J,J'=J)}
\end{equation}

Consider matrix elements $<\frac{1}{2},-\frac{1}{2}|\rho_{13}|-\frac{1}{2},\frac{1}{2}>$ and $<-\frac{1}{2},\frac{1}{2}|\rho_{13}|\frac{1}{2},-\frac{1}{2}>$. Getting from the former element to the latter element is equivalent to swapping primed and unprimed variables. But from \eq{eq:twoBodyAligned2} we see that $\rho_{13}(t)$ is invariant under primed and unprimed variable swapping; hence $<\frac{1}{2},-\frac{1}{2}|\rho_{13}|-\frac{1}{2},\frac{1}{2}>=<-\frac{1}{2},\frac{1}{2}|\rho_{13}|\frac{1}{2},-\frac{1}{2}>$. This tells us the diagonal terms in \eq{eq:alignedTerm} are zero.

Thus we have shown that the aligned term in the kinetic equation, $M \mbox{Tr}_3[V_{13},\rho_{13}]$, is zero.

\subsection{Initially Anti-Aligned Particles Contribute To Kinetic Equation}

The argument for initially aligned particles that showed that $\rho^{13}=\rho^{12}=\rho^{34}=\rho^{24}=\rho^{31}=\rho^{21}=\rho^{23}=\rho^{42}=0$ also hold for the case of initially anti-aligned particles. However careful consideration of \eq{eq:twoBodyDensityMatrix} reveals that it is not invariant under the swapping of primed and unprimed variables, therefore the two-body density matrices for initially anti-aligned particles is not symmetric.  From \eq{eq:alignedTerm} we see that a non-symmetric density matrix will populate the diagonal terms, thus the anti-aligned term of the kinetic equation will contribute to the kinetic equation.

Therefore the one-body kinetic equation reduces to,
\begin{equation}
\label{eq:kineticEq}
i \frac{\partial \rho_1}{\partial t}=N Tr_2[V_{12},\rho_{12}]
\end{equation}
where $N$ is the number of particles in opposite spin to particle 1, and particle 2 labels a particle that is initially in opposite spin to particle 1.

From now on, we will just refer to the one-body kinetic equation as the kinetic equation.

\subsection{Verification Of Consistency Of FMO Model With The Kinetic Equation}

As the kinetic equation is an exact equation, it stands to reason that the density matrices of the FMO model should satisfy the kinetic equation. In fact this would be important verification of the correctness of the one and two body density matrices of the FMO model. In this section we outline the proof that was used to show the consistency of the FMO density matrices with the kinetic equation.

The proof consists of two parts. In the first part, it is shown that the left and right-hand-side are of the same form,
\begin{equation}
i \frac{\partial \rho_1}{\partial t}=
\begin{pmatrix}
	A&0\\
	0&-A
\end{pmatrix}
\end{equation}
\begin{equation}
N Tr_n[V_{1n},\rho_{1n}]=
\begin{pmatrix}
	B&0\\
	0&-B
\end{pmatrix}
\end{equation}
In the second part through general angular momentum properties and physical constraints specific to the FMO model, it is shown that $A=B$. 

The complete proof of the consistency of the FMO density matrices with the kinetic equation can be found in Appendix \ref{ProofOfThe}.


\section{Analysis Of One-Body Factorisation On Kinetic Equation In FMO Model}
\label{AnalysisOneBodyFactorisation}

Our investigation into approximating the density matrix as a tensor product of one-body density matrices begins by seeing the effects it has on the kinetic equation in the FMO model. The kinetic equation with factorised density matrix becomes,
\begin{equation}
\label{eq:factorisedKineticEq}
i \frac{\partial \rho_1}{\partial t}=N Tr_2[V_{12},\rho_1\otimes\rho_2]
\end{equation}

\eq{eq:factorisedKineticEq} is in fact analogous to the classical Boltzmann equation which is the transport equation for dilute gases. Implicit in the derivation of the Boltzmann equation is the Boltzmann property,
\begin{equation}
\rho=\prod_{n=1}^N\rho_n
\end{equation}
the quantum analogue of which we are investigating. For brevity \eq{eq:factorisedKineticEq} will be referred to as the Boltzmann equation in the rest of this paper.

Let us start by revisiting the FMO one-body density matrix, which we present here in matrix element form,
\begin{equation}
\label{eq:oneBodyElem}
\begin{split}
		&<m_1|\rho_1|m_1'>\\
		&\quad=g\sum_{J,J',j,m}\exp\{-itg(J(J+1)-J'(J'+1))\}\\
		&\quad\times(2j_U+1)(2j+1)\\
		&\quad\times<j_U j_D; J M | m_U m_D> <j_U j_D; J' M | m_U m_D>\\
		&\quad\times<j_1 j; J M | m_1 m> <j_1 j; J' M | m_1' m>\\
		&\quad\times\left\{
			\begin{array}{ccc}
				1/2 & j_U-1/2 & j_U \\
				j_D & J & j
			\end{array}
		\right\}
		\left\{
			\begin{array}{ccc}
				1/2 & j_U-1/2 & j_U \\
				j_D & J' & j
			\end{array}
		\right\}
\end{split}
\end{equation}

Now the term $<j_1 j; J M | m_1 m> <j_1 j; J' M | m_1' m>$ forces the condition $M=m_1+m=m_1'+m$, which implies that the only density matrix element that is non-zero is when $m_1=m_1'$ i.e. the FMO one-body density matrix is diagonal. Therefore the tensor product of the one-body density matrices i.e $\rho_1\otimes\rho_n$ is also diagonal.

Now if we recall \eq{eq:alignedTerm}), which says that for a general two-body density matrix,
\begin{equation}
{Tr}_n[V_{1n},\rho_{1n}]=g
\begin{pmatrix}
	\rho^{32}-\rho^{23}&\rho^{13}-\rho^{12}+\rho^{34}-\rho^{24}\\
	\rho^{21}-\rho^{31}+\rho^{42}-\rho^{43}&\rho^{23}-\rho^{32}
\end{pmatrix}
\end{equation}
then,
\begin{equation}
N \mbox{Tr}_2[V_{12},\rho_1\otimes\rho_2]=0
\end{equation}
as $\rho_1\otimes\rho_2$ is a diagonal matrix.

This was a result we did not expect. Of course in general $i\frac{\partial \rho}{\partial t}$ is not zero. So what does this all mean?

To answer this question we need to realise that the density matrix can be written as the sum of the tensor product of the one-body density matrix and the correlation matrix\footnote, 
\begin{equation}
\rho=\prod_{n}\otimes\rho_{n}+\Gamma 
\end{equation}
So if we rewrite the kinetic equation with the correlation matrix explicitly included, 
\begin{equation}
\begin{split}
\label{factorisedKineticEqwithCorrelation}
i \frac{\partial \rho_1}{\partial t}&=N\mbox{Tr}_n[V_{1n},\rho_{1n}]\\
&=N\mbox{Tr}_n[V_{1n},\rho_1\otimes\rho_n]+N\mbox{Tr}_n[V_{1n},\Gamma]\\
&=N\mbox{Tr}_n[V_{1n},\Gamma]
\end{split}
\end{equation}
we clearly see that the evolution of the system in the FMO model is entirely driven by the correlation matrix.

Let's now take a moment and recall the works of Bell et. al \cite{bell03} and Friedland et al.\cite{friedland03a}\cite{friedland06}. The premise in these works was that the probability of flavour change due to coherent scattering is,
\begin{equation}
P(\nu_{\alpha}\rightarrow\nu_{\beta})=\left\lvert \sum_iA_i\right\rvert^2=R^2|A|^2
\end{equation}
and the probability of flavour change due to incoherent scattering is,
\begin{equation}
P(\nu_{\alpha}\rightarrow\nu_{\beta})=\sum_i|A_i|^2=R|A|^2
\end{equation}
where $A$ is the amplitude of a scattering event occurring and $R$ is the number of background neutrinos.

The time scales on which the probability evolves is,
\begin{align}
t_{coh}\propto R^{-1}\\
t_{inc}\propto R^{-\frac{1}{2}}
\end{align}
for coherent and incoherent scattering respectively. See \cite{friedland03a}\cite{friedland06}\cite{bell03} for further details.

Hence if the system exhibited flavour conversion in timescales which was not of an incoherent nature, then this was an indication of the breakdown of the one-body approximation, as the one-body approximation predicts conversion in incoherent timescales. Friedland found that in the FMO model, flavour evolution was of an incoherent nature, citing evidence of no breakdown in the one-body description. However what we have found here is that the flavour evolution of the FMO model is entirely driven by the correlation matrix, and in fact the one-body description would describe no flavour evolution at all; clearly there \emph{is} a breakdown in the one-body description in the FMO model.

\subsection{Effects Of Adding Neutrino Vacuum Oscillations}

\begin{figure}
\centering
  \includegraphics{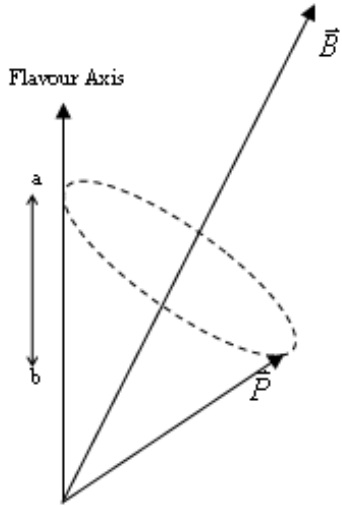}
\caption{$\vec{B}$ is the magnetic field vector about which spin vector $\vec{P}$ precesses. The spin projection on to the flavour axis determines the probability of finding a neutrino in that flavour. The value of this probability will oscillate between $a$ and $b$, the maximum and minimum value of the projection of $\vec{P}$ onto the flavour axis.}
\label{spinPrecession}
\end{figure}

The standard FMO model only encompasses flavour evolution through neutrino-neutrino interaction. We wish to extend this model to include the neutrino vacuum oscillation. To achieve this, as originally suggested by Okuniewicz \cite{okuniewicz06}, we introduce a constant magnetic field to the system. This works because we are representing neutrino flavour as spin, and the introduction of the magnetic field will cause these spin vectors to precess about the magnetic field. As the spin vector precesses, the spin projection on to some axis which represents a flavour state will oscillate (as depicted in Fig.\ref{spinPrecession}). This means the probability of being in some flavour state will oscillate representing the phenomenon of neutrino vacuum oscillations.
The reason we are interested in adding neutrino vacuum oscillation to the model is that it will populate the off-diagonal terms in the one-body density matrices - remember it was because of the diagonality of the one-body density matrices in the FMO model that had lead the factorised term to zero.

The derivation of the density matrices with oscillations\footnote{In the literature \emph{neutrino oscillations} implicitly mean \emph{neutrino vacuum oscillation}. This is the terminology we will use also from now on.} follows in a similar fashion to the system without oscillations, except that now the Hamiltonian has an extra single particle term, 
\begin{align}
	{H}&={K}_i+{V}_{ij}\\
	&={\vec{B}\cdot}{\vec{J}}+g{J}^2+g\frac{3}{4}N(N-2)
\end{align}
where $\vec{B}$ is the magnetic field vector, designed to be constant.
The one and two body density matrices with oscillation can be found in Appendix \ref{oneTwoOscillations}.

Including oscillations, the kinetic equation is now
\begin{equation}
i \frac{\partial \rho_1}{\partial t}=K\rho_1+N Tr_2[V_{12},\rho_1\otimes\rho_2]
\end{equation}
With density matrices which are now in general not diagonal, the argument presented at the beginning of this section that shows that $Tr_2[V_{12},\rho_1\otimes\rho_2]$ is zero no longer holds. Surprisingly however, a numerical analysis will maintain that this term is zero. This tells us that, at least for the numerical cases we have tried, that irrespective of whether we include vacuum oscillation in our model, the factorised term is still zero. The next section will show that in fact, in general, extending the FMO model to include vacuum oscillation will not alter the actuality that the dynamics of the system is completely driven by correlation affects.


\section{Why The One-Body Factorisation Breaks Down In The FMO Model}
\label{WhyBreak}
To understand why the one-body description completely breaks down in the FMO model, we ask the converse question: when will the one-body description not completely breakdown?

To ascertain this, let's rewrite the factorised term in the following way,
\begin{widetext}
\begin{equation}
\begin{split}
Tr_2[V_{12},\rho_1\otimes\rho_2]&=\sum_{m_2}<m_1 m_2|[V_{12},\rho_1\otimes\rho_2|m_1'' m_2>\\
&=\sum_{m_2}<m_1 m_2|V_{12} \rho_1\otimes\rho_2-\rho_1\otimes\rho_2 V_{12}|m_1'' m_2>\\
&=\sum_{m_2,m_1',m_2'}<m_1 m_2|V_{12}|m_1' m_2'><m1'|\rho_1|m_1''><m_2'|\rho_2|m_2>\\
&\quad-<m_1|\rho_1|m_1'><m_2|\rho_2|m_2'><m_1' 	m_2'|V_{12}|m_1'' m_2>\\
&=\sum_{m_2,m_1',m_2'}<m_1 m_2|V_{12}|m_1' m_2'><m_2'|\rho_2|m_2><m1'|\rho_1|m_1''>\\
&\quad-<m_1|\rho_1|m_1'><m_1' m_2'|V_{12}|m_1'' m_2><m_2|\rho_2|m_2'>
\end{split}
\end{equation}
\end{widetext}

Let
\begin{equation}
<m_1|f_1|m_1'>=\sum_{m_2,m_2'}{<m_1' m_2'|V_{12}|m_1'' m_2><m_2|\rho_2|m_2'>}
\end{equation}
Then,
\begin{equation}
\begin{split}
Tr_2[V_{12},\rho_1\rho_2]&=\sum_{m_1'}<m_1|f_1|m_1'><m_1'|\rho_1|m_1''>-<m_1|\rho_1|m_1'><m_1'|f_1|m_1''>\\
&=<m_1|[f_1,\rho_1]|m_1'>
\end{split}
\end{equation}
Now recall from \S\ref{sec:Hamiltonian_FMO} that the interaction Hamiltonian between two particles is $\frac{g}{2}(3+\vec{\sigma_i}\cdot\vec{\sigma_j})$.

So if we write the density matrix as a linear combination of the Pauli matrix basis,
\begin{align}
\label{eq:rho1Pauli}
\rho_1=\frac{1}{2}(P_0+\vec{P}\cdot\vec{\sigma_1})\\
\label{eq:rho2Pauli}
\rho_2=\frac{1}{2}(Q_0+\vec{Q}\cdot\vec{\sigma_2})
\end{align}
and realise that the only term in the interaction Hamiltonian that doesn't necessary vanish in commutation is $\frac{g}{2}\vec{\sigma_1}\cdot\vec{\sigma_2}$, then $f_1$ can be expressed as,
\begin{equation}
\begin{split}
f_1&=Tr_2(V_{12} \rho_2)\\
&=\frac{g}{4}Tr_2(\sigma_{1l}\sigma_{2l}(Q_0+Q_m\sigma_{2m}))\\
&=\frac{g}{4}Tr_2(\sigma_{1l}\sigma_{2l} Q_0+\sigma_{1l}Q_m(\delta_{lm}+i\epsilon_{lmn}\sigma{2n}))\\
&=\frac{g}{4}Tr_2(\vec{\sigma_1}\cdot\vec{Q})\\
&=\frac{g}{2}\vec{\sigma_1}\cdot\vec{Q}
\end{split}
\end{equation}
Hence,
\begin{equation}
\label{eq:specialPrecessionTerm}
\begin{split}
Tr_2[V_{12},\rho_1\otimes\rho_2]&=\frac{g}{4}[\vec{\sigma}\cdot\vec{Q},P_0+\vec{\sigma_1}\cdot\vec{P}]\\
&=\frac{g}{4}Q_l P_m[\sigma_{1l},\sigma_{1m}]\\
&=i \frac{g}{2} \epsilon_{lmn} \sigma_{1n} Q_l P_m\\
&=i \frac{g}{2}(\vec{Q}\times\vec{P})\cdot\vec{\sigma_1}
\end{split}
\end{equation}
\eq{eq:specialPrecessionTerm} tells us that the evolution of the system will be driven entirely by the correlation matrix when the interacting spins are parallel, and conversely the factorised term will contribute to the dynamics of the system when the interacting spins are not parallel. Now we have found that for our model that the factorised term is zero for all time; this suggests that the interacting spins are always parallel.

Now the left-hand-side of kinetic equation can be written in the following way,
\begin{equation}
\label{eq:lhsKineticEq}
\begin{split}
i\frac{\partial\rho_1}{\partial t}&=\frac{1}{2}\frac{\partial\vec{P}}{\partial t}\cdot\vec{\sigma_1}
\end{split}
\end{equation}
Combining \eq{eq:specialPrecessionTerm},\eq{eq:lhsKineticEq} the Boltzmann equation becomes,
\begin{equation}
\label{eq:specialPrecessionEq}
\frac{\partial\vec{P}}{\partial t}=Ng(\vec{Q}\times\vec{P})
\end{equation}

\eq{eq:specialPrecessionEq} is of a familiar form; it is indeed the equation for precession. It tells us that that the time evolution of a self-interacting neutrino system can be exactly described by the factorisation into one-body states only if the interacting spins precess around one another. As the one-body description yields absolutely no dynamics, this clearly is not happening. Why this is so, is evident if we recall that our one-body density matrix in the FMO model is diagonal in the spin basis. What this means is that if you were to single out an individual particle, which you necessary do in the one-body description, you will only ever find it in the spin-up or spin-down state. Particles which are spin-up or spin-down are parallel and therefore their spin vectors can never precess around one another.

\subsection{Breakdown Of Oscillation Extended FMO Model}
\begin{figure}
\centering
  \includegraphics{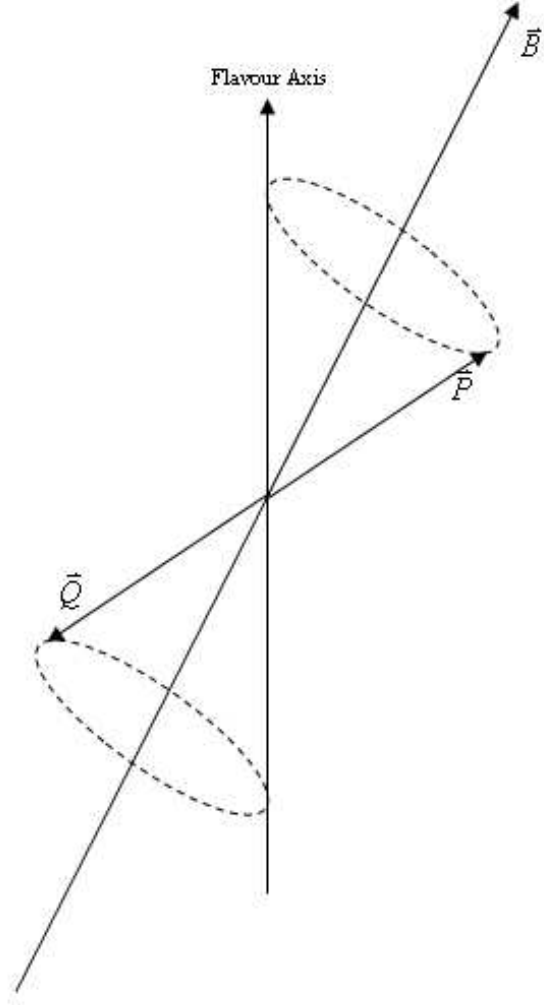}
\caption{Spin vector $\vec{P}$ began its life aligned with the flavour axis. $\vec{Q}$ began anti-aligned with the flavour axis. When $\vec{B}$ was switched on, $\vec{P}$ began to precess. Looking along the direction $\vec{P}$, it began precessing clockwise. Looking along the direction of $\vec{Q}$, $\vec{Q}$ also precesses, but in an anti-clockwise direction such that $\vec{P}$ and $\vec{Q}$ draw out two cones.}
\label{singleSpinPrecession}
\end{figure}

The above analysis also explains why the factorised term in the Oscillation Extended FMO model is zero. Initially in the FMO model the system starts in an untangled state with every particle in either an up or down spin state. At $t=0$ when the magnetic field is switched on, spin-up particles begin to precess, say, clock-wise around the magnetic field vector $\vec{B}$. Spin-down particles precess in an opposite direction, that is, anti-clockwise. We define (anti-)clockwise by looking along the direction of the vector that we are describing. As illustrated in Fig. \ref{singleSpinPrecession} the particles spin vectors will remain parallel drawing two cones; hence the introduction of the magnetic field, although causing the particles to precess does not distort them from their parallel configuration. Therefore by \eq{eq:specialPrecessionEq} the one-body description in the Oscillation Extended FMO Model also breaks down.


\section{How The Kinetic Equation Explains Flavour Evolution In FMO Model}

We extend the analysis in Chapter \ref{WhyBreak} to show how the kinetic equation can represent the complex flavour evolution in the FMO Model found in the work of Friedland et. al \cite{friedland06} and Okuniewicz\cite{okuniewicz06}.

The interaction Hamiltonian of the FMO model is a special case of a more general form, 
\begin{equation}
\label{eq:V12}
V_{12}=v_{ij}\sigma_{1i}\sigma_{2j}
\end{equation}
where $i,j=0..3$, and summation over repeated indices are implied.

We can also express the two-body density matrix in a similar fashion,
\begin{equation}
\label{eq:rho12}
\rho_{12}=p_{ij}\sigma_{1i}\sigma_{2j}
\end{equation}
Now,
\begin{equation}
\begin{split}
[V_{12},\rho_{12}]&=v_{ij}p_{kl}[\sigma_{1i}\sigma_{2j},\sigma_{1k}\sigma_{2l}]\\
&=v_{ij}p_{kl}(\sigma_{1i}\sigma_{2j}\sigma_{1k}\sigma_{2l}-\sigma_{1k}\sigma_{2l}\sigma_{1i}\sigma_{2j})\\
&=v_{ij}p_{kl}(\sigma_{1i}\sigma_{1k}\sigma_{2j}\sigma_{2l}-\sigma_{1i}\sigma_{1k}\sigma_{2j}\sigma_{2l}\\
&\quad+\sigma_{1i}\sigma_{1k}\sigma_{2j}\sigma_{2l}+\sigma_{1k}\sigma_{1i}\sigma_{2l}\sigma_{2j})\\
&=v_{ij}p_{kl}(\sigma_{1i}\sigma_{1k}[\sigma_{2j},\sigma_{2l}]+[\sigma_{1i},\sigma_{1k}]\sigma_{2l}\sigma_{2j})
\end{split}
\end{equation}
Since $\sigma_{10}$ and $\sigma_{20}$ are just the identity operators, they disappear in the commutation. Thus,
\begin{equation}
\begin{split}
[V_{12},\rho_{12}]=2iv_{ij}p_{kl}(\epsilon_{jlm}\sigma_{1i}\sigma_{1k}\sigma_{2m}+\epsilon_{ikn}\sigma_{1n}\sigma_{2l}\sigma{2j})
\end{split}
\end{equation}
where $i,k>0$ in the first term, and $j,l>0$ in the second term.

Further,
\begin{equation}
\label{eq:traceTerm1}
\begin{split}
\mbox{Tr}_2[V_{12},\rho_{12}]&=2iv_{ij}p_{kl}\mbox{Tr}_2(\epsilon_{jlm}\sigma_{1i}\sigma_{1k}\sigma{2m}+\epsilon_{ikn}\sigma_{1n}\sigma_{2l}\sigma{2j})\\
&=4iv_{ij}p_{kl}\epsilon_{ikn}\sigma_{1n}\delta_{lj}\\
&=4iv_{il}p_{kl}\epsilon_{ikn}\sigma_{1n}
\end{split}
\end{equation}
where $i,k>0$.

Before continuing let us make the following definitions,
\begin{align}
\label{eq:matrixVdef}
v&\equiv \mbox{matrix}\{v_{ij}\}\\
\label{matrixPdef}
p&\equiv \mbox{matrix}\{p_{ij}\}\\
\label{matrixWdef}
w&\equiv \mbox{matrix}\{v_{kl}\}\\
\label{matrixQdef}
q&\equiv \mbox{matrix}\{4p_{kl}\}
\end{align}
where $i,j=0..3$ and $k,l=1..3$.

The usefulness of these definitions will be apparent soon.

Now we know something about $p$. Firstly,
\begin{align}
\rho_1=\mbox{Tr}_2\rho_{12}=2p_{i0}\sigma_{1i}\\
\rho_2=\mbox{Tr}_1\rho_{12}=2p_{0j}\sigma_{2j}
\end{align}
So expressing $\rho_1$ and $\rho_2$ as a linear combination of the Pauli matrix basis as in \eq{eq:rho1Pauli} and \eq{eq:rho2Pauli} and applying the condition that the trace of density matrices equal 1, then we know that $p$ must at least be of the form,
\begin{equation}
\label{eq:pij}
p=\frac{1}{4}
\begin{pmatrix}
1&Q_1&Q_2&Q_3\\
P_1&p_{11}&p_{12}&p_{13}\\
P_2&p_{21}&p_{22}&p_{23}\\
P_3&p_{31}&p_{32}&p_{33}
\end{pmatrix}
\end{equation}
It is prudent to be clear of the distinguishment between \eq{eq:pij} and \eq{eq:rho1n4}. The latter is a density matrix, whereas the former is a co-efficient matrix. Their exact relation is described in \eq{eq:rho12}, where \eq{eq:rho1n4} is the $lhs$ and \eq{eq:pij} is the matrix co-efficient on the $rhs$. 

With this information about $p$ in hand, we can express \eq{eq:traceTerm1} as,
\begin{equation}
\label{eq:traceTerm2}
\begin{split}
\mbox{Tr}_2[V_{12},\rho_{12}]&=4iv_{i0}p_{k0}\epsilon_{ikn}\sigma_{1n}+4iv_{ij}p_{kj}\epsilon_{ikn}\sigma_{1n}\\
&=i(\vec{v_0}\times\vec{P})\cdot\sigma_1+i(wq^T)_{ik}\epsilon_{ikn}\sigma_{1n}\\
&=i(\vec{v_0}\times\vec{P})\cdot\sigma_1+i(wq^T)^D\cdot\sigma_1
\end{split}
\end{equation}
\begin{tabbing}
where \=$i,j,k,n=1..3$, and\\
\>$\vec{v_0}\equiv(v_{10},v_{20},v_{30})$
\end{tabbing}

In \eq{eq:traceTerm2} we have borrowed from exterior differential calculus the notion of a dual of a matrix i.e. for a general 3x3 matrix M, the vector $M_k^D\equiv\epsilon_{ijk}M_{ij}$.\footnote{In mathematical text the dual is usually represented with $^*$, e.g. $M_k^*$. We have chosen to use $^D$ instead to avoid confusion with the complex conjugate which is also sometimes represented with a $^*$.}

Therefore the kinetic equation can be written as,
\begin{equation}
\label{eq:exactPrecession}
\frac{1}{2N}\frac{\partial \vec{P}}{\partial t}=(\vec{v_0}\times\vec{P})+(wq^T)^D
\end{equation}

In the FMO model, 
\begin{equation}
V_{12}=\frac{3}{2}g+\frac{g}{2}\sigma_1\sigma_2
\end{equation}
hence,
\begin{equation}v^{FMO}=\frac{g}{2}
\label{eq:vijFMO}
\begin{pmatrix}
3&0&0&0\\
0&1&0&0\\
0&0&1&0\\
0&0&0&1
\end{pmatrix}
\end{equation}
Again we make clear the difference between the two matrices \eq{eq:vijFMO} and \eq{eq:Vij4}. The latter is the interaction Hamiltonian in matrix form, the former is a matrix co-efficient.  Their exact relation is described in \eq{eq:V12}, where \eq{eq:Vij4} is the $lhs$ and \eq{eq:vijFMO} is the matrix co-efficient on the $rhs$. 

From \eq{eq:vijFMO} we see that $\vec{v_0}=0$ and $w=\frac{g}{2}\dblone$. Therefore if we revisit \eq{eq:exactPrecession} with the FMO Hamiltonian, the first term disappears but we are left with the second term,
\begin{equation}
\label{eq:precessionqT}
\frac{\partial \vec{P}}{\partial t}=Ng(q^T)^D
\end{equation}
The probability of finding a neutrino in a certain flavour state is determined by the projection of $\vec{P}$ on to the axis that represents that flavour state. Hence \eq{eq:precessionqT} is exactly what drives the the flavour evolution in the FMO Model.

\eq{eq:precessionqT} is the exact kinetic equation of the FMO model. As a check, we should be able to retrieve the factorised kinetic equation by applying the Boltzmann property,
\begin{equation}
\label{eq:rho12Boltz}
\begin{split}
\rho_{12}&=\rho_1\otimes\rho_2\\
&=\frac{1}{4}P_iQ_j\sigma_{1i}\sigma_{2j}\\
\end{split}
\end{equation}
where $i,j=0..3$

Comparing \eq{eq:rho12Boltz} with \eq{eq:rho12} we see that $p_{ij}=\frac{1}{4}P_iQ_j$. Therefore,
\begin{equation}
\label{eq:qT}
\begin{split}
(q^T)^D&=\epsilon_{lmn}4p_{ml}\\
&=\epsilon_{lmn}P_mQ_l\\
&=\vec{Q}\times\vec{P}
\end{split}
\end{equation}
where $l,m,n=1..3$

Substitution of \eq{eq:qT} into \eq{eq:precessionqT} recovers exactly the factorised kinetic equation of \eq{eq:specialPrecessionEq}.

Furthermore, from \eq{eq:precessionqT} we able to reveal that the FMO is driven entirely by the correlation term. To see this we first expand out $(q^T)^D$,
\begin{equation}
\begin{split}
(q^T)^D&=\epsilon_{ijk}q_{ij}^T\\
&=(q_{23}^T-q_{32}^T,q_{31}^T-q_{13}^T,q_{12}^T-q_{21}^T)
\end{split}
\end{equation}
Now recall,
\begin{equation}
\label{eq:rhoCorr}
\rho_{12}=\rho_1\otimes\rho_2+\Gamma\\
\end{equation}
If we define the correlation matrix in the following way,
\begin{equation}
\Gamma\equiv\frac{1}{4}\gamma_{ij}\sigma_{1i}\sigma_{2j}\\
\end{equation}
then \eq{eq:rhoCorr} can be written as,
\begin{equation}
p_{ij}=\frac{1}{4}P_iQ_j+\frac{1}{4}\gamma_{ij}\\
\end{equation}
With the constraints imposed on $\gamma_{ij}$ by \eq{eq:pij}, we see that,
\begin{equation}
\label{eq:pijgamma}
p=\frac{1}{4}
\begin{pmatrix}
1&Q_1&Q_2&Q_3\\
P_1&P_1Q_1+\gamma_{11}&P_1Q_2+\gamma_{12}&P_1Q_2+\gamma_{13}\\
P_2&P_2Q_1+\gamma_{21}&P_2Q_2+\gamma_{22}&P_2Q_2+\gamma_{23}\\
P_3&P_3Q_1+\gamma_{31}&P_3Q_2+\gamma_{32}&P_3Q_2+\gamma_{33}\\
\end{pmatrix}
\end{equation}
Hence,
\begin{equation}
\begin{split}
(q^T)^D&=(P_3Q_2+\gamma_{32}-P_2Q_3+\gamma_{23},P_1Q_3+\gamma_{13}-P_3Q_1+\gamma_{31},P_2Q_1+\gamma_{21}-P_1Q_2+\gamma_{12})\\
&=\vec{Q}\times\vec{P}+(\gamma^T)^D
\end{split}
\end{equation}
Now we know that in the FMO model $\vec{Q}\times\vec{P} = 0$, thus,
\begin{equation}
\label{eq:precessiongammaT}
\frac{\partial \vec{P}}{\partial t}=Ng(\gamma^T)^D
\end{equation}
Therefore in \eq{eq:precessiongammaT} we have shown how the flavour evolution in the FMO model is entirely driven by the correlations.


\section{Conclusions}
\label{ChConclusion}

The ability of physics to abstract complex systems to simple ones by stripping away that which is not essential has been extremely successful; it has allowed mankind to comprehend his Universe to levels of detail only dreamt about in the past. However the strength of these simplified models ironically rests on the validity of what is not included in the model. Time and time again in physics' past, questioning the legitimacy of these assumptions have opened a Pandora's box of insights and new physics.

To describe the kinetic equations of dense neutrino systems such as early Universe and Supernovae, the literature has assumed the Boltzmann property where many-body states can be portrayed as the product of the individual states of the system. Debate over the validity of this assumption in such systems has recently surfaced. A technique which has been used to test the legitimacy of this assumption is to compare the neutrino flavour evolution of the exact many-body systems to that which is predicted by the product of one-body states. Since the one-body description predicts flavour evolution to be of an incoherent timescale, if the exact system was found to have flavour evolution which was not of an incoherent nature, than this would signal the breakdown in the one-body approximation. The Friedland-McKellar-Okuniewicz model was developed to conduct this test. It was found by the authors of the model that their system evolved in incoherent timescales, and hence they concluded support for the one-body approximation. Our investigations found that the dynamics of the FMO model was entirely driven by the correlation function, and in fact the one-body description would describe no evolution in the system at all; there clearly was a breakdown in the one-body description in the FMO model.

It is important to recognise that the Boltzmann factorisation of the two-body density matrix was not the only assumption used in the kinetic equations of the early Universe and Supernovae.  Another assumption central to the derivation of the neutrino kinetic equations was that the duration of a collision was short compared to the time between collisions.  The same assumption was made in Boltzmann's derivation of his equation for the kinetics of dilute gases. In the FMO model there is not spatial dependence in the interactions, so that the interactions are never turned off. The duration of the interactions is thus the whole time elapsed.  With this in mind, perhaps we should not be so surprised that the Boltzmann-type of kinetic equation breaks down in the FMO model.  Notwithstanding the increased complexity, a better model to test the Boltzmann property with will need to include or at least mimic a spatial extent.

Our conclusion that the one-body description breaks down in the FMO model is in stark contrast to the result found by Friedland et. al. This questions the effectiveness of using neutrino flavour timescales to test the validity of the one-body approximation. Even though the FMO model evolved in the same timescales as that predicted by the one-body approximation, this is not an indication of the validity of the approximation, as we have shown that this approximation in the FMO model would catastrophically breakdown. We conclude that at best, the technique of comparing neutrino flavour evolutionary timescales can only invalidate the Boltzmann property but never validate it.

\section{Acknowledgments}

Thank you to Bruce H.J. McKellar for his kind supervision.

\appendix

\section{Proof Of The Consistency Of The Density Matrices With The Kinetic Equation}
\label{ProofOfThe}

In this appendix we show that the one-body density matrix and two-body density matrix derived in \S\ref{sec:One_Two_Density_Matrices_FMO} is consistent with the kinetic equation \eq{eq:kineticEq}. The proof is divided up into logical sections \ref{lhsForm}-\ref{showEquivlhsrhs}.
\section{Form Of Left-Hand-Side}
\label{lhsForm}
Let's begin by differentiating \eq{eq:oneBodyDensityMatrixUp} to get the left-hand-side of the kinetic equation (note although we are considering the case when the particle's initial configuration is spin-up, it also holds for the case when the particle is spin down).
\begin{equation}
\label{eq:oneBodyDensityElem}
\begin{split}
		i\frac{\partial\rho_1}{\partial t}&=g\sum_{J,J',j,m}(-)^{(J-J')}(J(J+1)-J'(J'+1))/exp\{-itg(J(J+1)-J'(J'+1))\}(2j_U+1)(2j+1)\\
		&\quad<j_U j_D; J M | m_U m_D> <j_U j_D; J' M | m_U m_D><j_1 j; J M | m_1^{row} m> <j_1 j; J' M | m_1^{col} m>\\
		&\quad\left\{
			\begin{array}{ccc}
				1/2 & j_U-1/2 & j_U \\
				j_D & J & j
			\end{array}
		\right\}
		\left\{
			\begin{array}{ccc}
				1/2 & j_U-1/2 & j_U \\
				j_D & J' & j
			\end{array}
		\right\}
\end{split}
\end{equation}
Here we have expressed the $lhs$ in matrix element form. $m_1^{row}$ and $m_1^{col}$ represent the spin projection basis of the row and column of the matrix. One sees that the condition $M=m_1^{row}+m=m_1^{col}+m$ in \eq{eq:oneBodyDensityElem} implies that the $lhs$ is diagonal in this basis.

Next we acknowledge the following two properties:
\begin{itemize}
\item that for $J=J'$, the terms in the summation go to zero since $J(J+1)-J'(J'+1)=0$
\item from the triangular constraints in the 6j-symbol, $J,J'\in\{|j-\frac{1}{2}|,j+\frac{1}{2}\}$
\end{itemize}

If we combine these properties with the known Clebsch-Gordan co-efficient relation,
\begin{equation}
<j_1j_2m_1m_2|j_1j_2JM>=(-)^{J-j_1-j_2}<j_2j_1m_2m_1|j_2j_1JM>
\end{equation}
we see that $i\frac{\partial\rho_1^{(\frac{1}{2},\frac{1}{2})}}{\partial t} =-i\frac{\partial\rho_1^{(-\frac{1}{2},-\frac{1}{2})}}{\partial t}$, where the superscript is the $(m_1^{row},m_1^{col})$ of \eq{eq:oneBodyDensityElem}, which identifies the matrix element.

Let's now have a look at the $rhs$ of the kinetic equation.

\subsection{Form Of Right-Hand-Side}
\label{rhsForm}

The interaction Hamiltonian in matrix form is,
\begin{equation}
\label{eq:Vij2}
V_{ij}=\frac{3g}{2}\dblone+\frac{g}{2}
\begin{pmatrix}
1&0&0&0\\
0&-1&2&0\\
0&2&-1&0\\
0&0&0&1
\end{pmatrix}
\end{equation}
where $\dblone$ is the unit matrix.The spin basis of \eq{eq:Vij2} reading left to right (and of course top to bottom) is, $|\uparrow\uparrow>$,$|\uparrow\downarrow>$,$|\downarrow\uparrow>$,$|\downarrow\downarrow>$, where the first arrow represents the spin of particle one, and the second the spin of particle two.

Let us now define a general two-body density matrix,
\begin{equation}
\label{eq:rho1n2}
\rho_{12}=
\begin{pmatrix}
\rho^{11}&\rho^{12}&\rho^{13}&\rho^{14}\\
\rho^{21}&\rho^{22}&\rho^{23}&\rho^{24}\\
\rho^{31}&\rho^{32}&\rho^{33}&\rho^{34}\\
\rho^{41}&\rho^{42}&\rho^{43}&\rho^{44}\\
\end{pmatrix}
\end{equation}
where we have omitted the subscripts in the matrix elements for clarity. \eq{eq:rho1n2} has the same basis as \eq{eq:Vij2}.
Using \eq{eq:Vij2} and \eq{eq:rho1n2} to evaluate the $rhs$ of the kinetic equation we get,
\begin{equation}
\label{eq:rhsKineticEqA}
N \mbox{Tr}_2[V_{12},\rho_{12}]=Ng
\begin{pmatrix}
	\rho^{32}-\rho^{23}&\rho^{13}-\rho^{12}+\rho^{34}-\rho^{24}\\
	\rho^{21}-\rho^{31}+\rho^{42}-\rho^{43}&\rho^{23}-\rho^{32}
\end{pmatrix}
\end{equation}
Now $m_{12}+m=M$. So for element $<\frac{1}{2},\frac{1}{2}|\rho_{12}|-\frac{1}{2},\frac{1}{2}>$,
\begin{equation}
\begin{split}
	<j_{1n} j; m_{1n} m |J M> <j'_{1n} j; m'_{1n} m |J' M>&=<1 j; 1 m |J M> <1 j; 0 m |J' M>\\
	&=0
\end{split}
\end{equation}
since it is not possible for $M$ to satisfy both $m_{12}+m=1+m$ and $m_{12}'+m=0+m$.

By same reasoning matrix elements, $\rho^{13}=\rho^{12}=\rho^{34}=\rho^{24}=\rho^{31}=\rho^{21}=\rho^{23}=\rho^{42}=0$.

Using this result, the off-diagonal terms of the $rhs$ matrix \eq{eq:rhsKineticEqA} are zero. As for the diagonal elements, \eq{eq:rhsKineticEqA} clearly tell us they are the negative of each other.

To this point we have shown that the $lhs$ and $rhs$ of the kinetic equation exhibit the same form, namely the off-diagonal matrix elements are zero, and the two diagonal elements are the negative of each other. The only thing left to do now is to show that one of the diagonal elements of the $lhs$ is equal to the equivalent one of the $rhs$.

\subsection{Showing The Equivalence Of $lhs$ and $rhs$ Diagonal Elements}
\label{showEquivlhsrhs}

We are going to show that $lhs^{(\frac{1}{2},\frac{1}{2})}=rhs^{(\frac{1}{2},\frac{1}{2})}$, where the superscripts have their usual meaning, in this case identifying the first element of the matrices.

\subsection{$lhs^{(\frac{1}{2},\frac{1}{2})}$}

Define,
\begin{equation}
\label{eq:alphaDef}
	\alpha(J,J')\equiv J(J+1)-J'(J'+1)
\end{equation}
\begin{equation}
\begin{split}
\label{betaDef}
	\beta(J,J')&\equiv (2j_U+1)(2j+1)\\
	&\quad<j_U j_D; J M | m_U m_D> <j_U j_D; J' M | m_U m_D><\frac{1}{2} j; J M | \frac{1}{2} m> <\frac{1}{2} j; J' M | \frac{1}{2} m>\\
		&\quad\left\{
			\begin{array}{ccc}
				1/2 & j_U-1/2 & j_U \\
				j_D & J & j
			\end{array}
		\right\}
		\left\{
			\begin{array}{ccc}
				1/2 & j_U-1/2 & j_U \\
				j_D & J' & j
			\end{array}
		\right\}
\end{split}
\end{equation}
Then,
\begin{equation}
\begin{split}
lhs^{(\frac{1}{2},\frac{1}{2})}&=\sum_{J,J',j}\alpha(J,J')(\cos(tg\alpha(J,J'))-i\sin(tg\alpha(J,J')))\beta(J,J')\\
&=\sum_{\substack{J<J',\\j}}-2i\alpha(J,J')\sin(tg\alpha(J,J')))\beta(J,J')\\
\end{split}
\end{equation}
The 6j-symbol in $\beta$ restricts $J,J'\in\{|j-\frac{1}{2}|,j+\frac{1}{2}\}$; and because $J<J'$,then $J=|j-\frac{1}{2}|,J'=j+\frac{1}{2}$. Note that if $J=J'$ then $\alpha=0$. With these substitution for $J,J'$ we get,
\begin{equation}
\label{eq:lhs1}
\begin{split}
lhs^{(\frac{1}{2},\frac{1}{2})}&=i2g(2j_U+1)\sum_{j=|j_U-\frac{1}{2}-j_D|}^{|j_U-\frac{1}{2}+j_D|}(2j+1)^2\sin(tg(2j+1))\\
&\quad<j_U j_D; j-\frac{1}{2} M | m_U m_D> <j_U j_D; j+\frac{1}{2} M | m_U m_D>\\
&\quad<\frac{1}{2} j; \frac{1}{2} M-\frac{1}{2} |j-\frac{1}{2} M><\frac{1}{2} j; \frac{1}{2} M-\frac{1}{2} |j+\frac{1}{2} M>\\
		&\quad\left\{
			\begin{array}{ccc}
				1/2 & j_U-1/2 & j_U \\
				j_D & j-\frac{1}{2} & j
			\end{array}
		\right\}
		\left\{
			\begin{array}{ccc}
				1/2 & j_U-1/2 & j_U \\
				j_D & j+\frac{1}{2} & j
			\end{array}
		\right\}\\
&=i2g(2j_U+1)\sum_{j=|j_U-j_D|}^{|j_U+j_D|}(2j)^2\sin(tg2j)\\
&\quad<j_U j_D; j-1 M | m_U m_D> <j_U j_D; j M | m_U m_D>\\
&\quad<\frac{1}{2} j-\frac{1}{2}; \frac{1}{2} M-\frac{1}{2} |j-1 M><\frac{1}{2} j-\frac{1}{2}; \frac{1}{2} M-\frac{1}{2} |j M>\\
		&\quad\left\{
			\begin{array}{ccc}
				1/2 & j_U-1/2 & j_U \\
				j_D & j-1 & j-\frac{1}{2}
			\end{array}
		\right\}
		\left\{
			\begin{array}{ccc}
				1/2 & j_U-1/2 & j_U \\
				j_D & j & j-\frac{1}{2}
			\end{array}
		\right\}
\end{split}
\end{equation}
In the last line of \eq{eq:lhs1}, we have just done a shift of the summed variable $j$ i.e. $j\rightarrow j-\frac{1}{2}$.

\subsubsection{$rhs^{(\frac{1}{2},\frac{1}{2})}$}\

\begin{equation}
rhs=Ng
\begin{pmatrix}
	\rho^{32}-\rho^{23}&0\\
	0&\rho^{23}-\rho^{32}
\end{pmatrix}
\end{equation}
Using the two-body density matrix \eq{eq:twoBodyDensityMatrix},
\begin{equation}
\label{eq:rhsA}
\begin{split}
rhs^{(\frac{1}{2},\frac{1}{2})}&=N\sum_{J,J',j}-\sqrt{3}A(J,J')(2j_D+1)<1 j; 0 m |J' M>\\
		&\quad\left\{
			\begin{array}{ccc}
				\frac{1}{2} & j_U-\frac{1}{2} & j_U \\
				\frac{1}{2} & j_D-\frac{1}{2} & j_D\\
				0 & j & J\\
			\end{array}
		\right\}
		\left\{
			\begin{array}{ccc}
				\frac{1}{2} & j_U-\frac{1}{2} & j_U \\
				\frac{1}{2} & j_D-\frac{1}{2} & j_D\\
				1 & j & J'\\
			\end{array}
		\right\}\\
		&\quad+\sqrt{3}A(J,J')(2j_D+1)<1 j; 0 m |J M>\\
		&\quad\left\{
			\begin{array}{ccc}
				\frac{1}{2} & j_U-\frac{1}{2} & j_U \\
				\frac{1}{2} & j_D-\frac{1}{2} & j_D\\
				0 & j & J'\\
			\end{array}
		\right\}
		\left\{
			\begin{array}{ccc}
				\frac{1}{2} & j_U-\frac{1}{2} & j_U \\
				\frac{1}{2} & j_D-\frac{1}{2} & j_D\\
				1 & j & J\\
			\end{array}
		\right\}\\
		&=N\sqrt{3}(2j_D+1)\sum_{J,J',j}<1 j; 0 m |J M>	\left\{
			\begin{array}{ccc}
				\frac{1}{2} & j_U-\frac{1}{2} & j_U \\
				\frac{1}{2} & j_D-\frac{1}{2} & j_D\\
				0 & j & j\\
			\end{array}
		\right\}\\
		&\quad(A(J,j)\left\{
			\begin{array}{ccc}
				\frac{1}{2} & j_U-\frac{1}{2} & j_U \\
				\frac{1}{2} & j_D-\frac{1}{2} & j_D\\
				1 & j & J\\
			\end{array}
		\right\}-A(j,J')\left\{
			\begin{array}{ccc}
				\frac{1}{2} & j_U-\frac{1}{2} & j_U \\
				\frac{1}{2} & j_D-\frac{1}{2} & j_D\\
				1 & j & J\\
			\end{array}
		\right\})
\end{split}
\end{equation}
where
\begin{equation}
\begin{split}
A(J,J')&=g\exp\{-itg(J(J+1)-J'(J'+1))\}(2j_U+1)(2j+1)\\
&\quad<j_Uj_D;JM|m_um_D><j_Uj_D;J'M|m_um_D>
\end{split}
\end{equation}
As $J$ and $J'$ range over the same values, by careful rearrangement of \eq{eq:rhsA} we can reduce $rhs^{(\frac{1}{2},\frac{1}{2})}$ to just a summation over $J,j$.
\begin{equation}
\begin{split}
rhs^{(\frac{1}{2},\frac{1}{2})}&=N\sqrt{3}(2j_D+1)\sum_{J,j}<1 j; 0 m |j M>
		\left\{
			\begin{array}{ccc}
				\frac{1}{2} & j_U-\frac{1}{2} & j_U \\
				\frac{1}{2} & j_D-\frac{1}{2} & j_D\\
				0 & j & j\\
			\end{array}
		\right\}
		\left\{
			\begin{array}{ccc}
				\frac{1}{2} & j_U-\frac{1}{2} & j_U \\
				\frac{1}{2} & j_D-\frac{1}{2} & j_D\\
				1 & j & J'\\
			\end{array}
		\right\}\\
		&\quad(A(J,j)-A(j,J))\\
		&=-N\sqrt{3}g(2j_D+1)(2j_U+1)2i\sum_{J,j}\sin(tg(J(J+1)-j(j+1))(2j+1)\\
		&\quad<j_Uj_D;JM|m_um_D><j_Uj_D;J'M|m_um_D><1 j; 0 m |J' M>\\
		&\quad\left\{
			\begin{array}{ccc}
				\frac{1}{2} & j_U-\frac{1}{2} & j_U \\
				\frac{1}{2} & j_D-\frac{1}{2} & j_D\\
				0 & j & j\\
			\end{array}
		\right\}
		\left\{
			\begin{array}{ccc}
				\frac{1}{2} & j_U-\frac{1}{2} & j_U \\
				\frac{1}{2} & j_D-\frac{1}{2} & j_D\\
				1 & j & J\\
			\end{array}
		\right\}
\end{split}
\end{equation}
The 9j-symbol restricts $J\in\{|j-1|,j+1\}$. With these substitutions for $J$ and the help of \eq{eq:special09J} and \eq	{eq:special19J} we get,
\begin{equation}
\label{eq:rhs1}
\begin{split}
rhs^{(\frac{1}{2},\frac{1}{2})}&=i4\sqrt{3}Ng(2j_U+1)(2j_D+1)\\
&\quad\sum_{j=|j_U-j_D|}^{|j_U+j_D-1|}(2j+1)<j_Uj_D;jM|m_Um_D>\\
&\quad\frac{(-)^{j_U+j_D+3j}}{\sqrt{2(2j+1)}}
		\left\{
			\begin{array}{ccc}
				j_U-\frac{1}{2} & j_U & \frac{1}{2} \\
				j_D & j_D-\frac{1}{2} & j
			\end{array}
		\right\}\\
		&\quad(\sin(tg2(2j+1))<j_Uj_D;j+1 M|m_Um_D><1j;0M|j+1 M>\\
		&\qquad((j+1)\{\mathcal{A'}\}\{\mathcal{B'}\}\{\mathcal{C'}\}+(j+2)\{\mathcal{D'}\}\{\mathcal{E'}\}\{\mathcal{F'}\})\\
		&\quad+\sin(-2jtg))<j_Uj_D;j-1 M|m_Um_D><1j;0M|j-1 M>\\
		&\qquad(j\{\mathcal{G}\}\{\mathcal{H}\}\{\mathcal{I}\}+(j-1)\{\mathcal{J}\}\{\mathcal{K}\}\{\mathcal{L}\}))
\end{split}
\end{equation}
where
\begin{equation}
\begin{split}
\{\mathcal{A'}\}&=	
		\left\{
			\begin{array}{ccc}
				\frac{1}{2} & j_U-\frac{1}{2} & j_U \\
				j_D & j+1 & j+\frac{1}{2}
			\end{array}
		\right\}\qquad
\{\mathcal{B'}\}=
		\left\{
			\begin{array}{ccc}
				\frac{1}{2} & j_D-\frac{1}{2} & j_D \\
				j_U-\frac{1}{2} & j+\frac{1}{2} & j
			\end{array}
		\right\}\qquad
\{\mathcal{C'}\}=
		\left\{
			\begin{array}{ccc}
				1 & j & j+1 \\
				j+\frac{1}{2} & \frac{1}{2} & \frac{1}{2}
			\end{array}
		\right\}\\
\{\mathcal{D'}\}&=
		\left\{
			\begin{array}{ccc}
				\frac{1}{2} & j_U-\frac{1}{2} & j_U \\
				j_D & j+1 & j+\frac{3}{2}
			\end{array}
		\right\}\qquad
\{\mathcal{E'}\}=
		\left\{
			\begin{array}{ccc}
				\frac{1}{2} & j_D-\frac{1}{2} & j_D \\
				j_U-\frac{1}{2} & j+\frac{3}{2} & j
			\end{array}
		\right\}\qquad
\{\mathcal{F'}\}=
		\left\{
			\begin{array}{ccc}
				1 & j & j+1 \\
				j+\frac{3}{2} & \frac{1}{2} & \frac{1}{2}
			\end{array}
		\right\}\\
\{\mathcal{G}\}&=
		\left\{
			\begin{array}{ccc}
				\frac{1}{2} & j_U-\frac{1}{2} & j_U \\
				j_D & j+1 & j+\frac{1}{2}
			\end{array}
		\right\}\qquad
\{\mathcal{H}\}=	
		\left\{
			\begin{array}{ccc}
				\frac{1}{2} & j_D-\frac{1}{2} & j_D \\
				j_U-\frac{1}{2} & j+\frac{1}{2} & j
			\end{array}
		\right\}\qquad
\{\mathcal{I}\}=
		\left\{
			\begin{array}{ccc}
				1 & j & j+1 \\
				j+\frac{1}{2} & \frac{1}{2} & \frac{1}{2}
			\end{array}
		\right\}\\
\{\mathcal{J}\}&=
		\left\{
			\begin{array}{ccc}
				\frac{1}{2} & j_U-\frac{1}{2} & j_U \\
				j_D & j+1 & j+\frac{3}{2}
			\end{array}
		\right\}\qquad
\{\mathcal{K}\}=	
		\left\{
			\begin{array}{ccc}
				\frac{1}{2} & j_D-\frac{1}{2} & j_D \\
				j_U-\frac{1}{2} & j+\frac{3}{2} & j
			\end{array}
		\right\}\qquad
\{\mathcal{L}\}=	
		\left\{
			\begin{array}{ccc}
				1 & j & j+1 \\
				j+\frac{3}{2} & \frac{1}{2} & \frac{1}{2}
			\end{array}
		\right\}\\
\end{split}
\end{equation}

It can be shown that \eq{eq:rhs1} is equivalent to,
\begin{equation}
\label{eq:rhs2}
\begin{split}
rhs^{(\frac{1}{2},\frac{1}{2})}&=i2\sqrt{6}Ng(2j_U+1)(2j_D+1)\\
&\quad\sum_{j=|j_U-j_D|}^{j_U+j_D}(-)^{j_U+j_D+3j+1}\sin(tg2j)<j_Uj_D;jM|m_Um_D><j_Uj_D;j-1M|m_Um_D>\\
&\quad\sqrt{2j-1}
		\left\{
			\begin{array}{ccc}
				j_U-\frac{1}{2} & j_U & \frac{1}{2} \\
				j_D & j_D-\frac{1}{2} & j-1
			\end{array}
		\right\}<1j-1;0M|jM>\\
		&\qquad(j\{\mathcal{A}\}\{\mathcal{B}\}\{\mathcal{C}\}+(j+1)\{\mathcal{D}\}\{\mathcal{E}\}\{\mathcal{F}\})\\
&\quad\sqrt{2j+1}
		\left\{
			\begin{array}{ccc}
				j_U-\frac{1}{2} & j_U & \frac{1}{2} \\
				j_D & j_D-\frac{1}{2} & j
			\end{array}
		\right\}<1j;0M|j-1M>\\
		&\qquad(j\{\mathcal{G}\}\{\mathcal{H}\}\{\mathcal{I}\}+(j-1)\{\mathcal{J}\}\{\mathcal{K}\}\{\mathcal{L}\})\\
\end{split}
\end{equation}
where,
\begin{equation}
\begin{split}
\{\mathcal{A}\}&=	
		\left\{
			\begin{array}{ccc}
				\frac{1}{2} & j_U-\frac{1}{2} & j_U \\
				j_D & j & j-\frac{1}{2}
			\end{array}
		\right\}\qquad
\{\mathcal{B}\}=
		\left\{
			\begin{array}{ccc}
				\frac{1}{2} & j_D-\frac{1}{2} & j_D \\
				j_U-\frac{1}{2} & j-\frac{1}{2} & j-1
			\end{array}
		\right\}\qquad
\{\mathcal{C}\}=
		\left\{
			\begin{array}{ccc}
				1 & j-1 & j \\
				j-\frac{1}{2} & \frac{1}{2} & \frac{1}{2}
			\end{array}
		\right\}\\
\{\mathcal{D}\}&=
		\left\{
			\begin{array}{ccc}
				\frac{1}{2} & j_U-\frac{1}{2} & j_U \\
				j_D & j & j+\frac{1}{2}
			\end{array}
		\right\}\qquad
\{\mathcal{E}\}=
		\left\{
			\begin{array}{ccc}
				\frac{1}{2} & j_D-\frac{1}{2} & j_D \\
				j_U-\frac{1}{2} & j+\frac{1}{2} & j
			\end{array}
		\right\}\qquad
\{\mathcal{F}\}=
		\left\{
			\begin{array}{ccc}
				1 & j-1 & j \\
				j+\frac{1}{2} & \frac{1}{2} & \frac{1}{2}
			\end{array}
		\right\}\\
\end{split}
\end{equation}
The important thing about \eq{eq:rhs2}, is that it has exactly the same summation limits as $lhs^{(\frac{1}{2},\frac{1}{2})}$ \eq{eq:lhs1}. So if we define,
\begin{equation}
lhs^{(\frac{1}{2},\frac{1}{2})}\equiv\sum_{j=|j_U-j_D|}^{j_U+j_D}X
\end{equation}
\begin{equation}
rhs^{(\frac{1}{2},\frac{1}{2})}\equiv\sum_{j=|j_U-j_D|}^{j_U+j_D}Y
\end{equation}
then with the help of \eq{eq:W3}-\eq{eq:3j3} and after some simple but tedious algebra, one arrives at,
\begin{equation}
\label{eq:XminusY}
\begin{split}
X-Y&=\frac{(-)^{-2(2j+j_U+j_D)}((-)^{2j}-(-)^{2M})(j-j_U-j_D)}{2jj_U(2j_U+1)}\\
&\quad \cdot\sqrt{\frac{(j+j_D-jU)(j-j_D+jU)(1-j+j_D+jU)(1+j+j_D+jU)(j-M)(j-M)}{(2j-1)(2j+1)}}
\end{split}
\end{equation}
Now if we impose the physical constraints,
\begin{align}
j&=j_U-J_D+c\\
M&=j_U-j_D
\end{align}
where $c$ is a positive integer, then,
\begin{equation}
\begin{split}
	(-)^{2j}-(-)^{2M}&=(-)^{2(j_U-J_D+c)}-(-)^{2(j_U-j_D)}\\
	&=(-)^{2(j_U-J_D)}((-)^{{2c}}-1)\\
	&=0
\end{split}
\end{equation}
Hence,
\begin{equation}
X-Y=0
\end{equation}
Thus we have shown the final piece of the proof, that $lhs^{(\frac{1}{2},\frac{1}{2})}=rhs^{(\frac{1}{2},\frac{1}{2})}$. Therefore we have proven the consistency of the density matrix with the kinetic equation.

Q.E.D.


\section{One And Two Body Density Matrices With Interaction and Vacuum Oscillation}
\label{oneTwoOscillations}

\begin{equation}
	\begin{split}
			\rho_1(t) &= \sum_{\substack{J,J',M^B,{M^B}'\\j,m^B,m_1^B,{m_1^B}'}}(-)^{J-J'} \exp\{-it[B(M^B-{M^B}')+g(J(J+1)-J'(J'+1))]\}\\
			&\quad d^J_{M^B,M}(-\beta)d^{J'}_{M',{M^B}'}(-\beta)d^{j_1}_{m_1,m_1^B}(-\beta)d^{j_1}_{{m_1^B}',{m_1}'}(\beta')(2j_U+1)(2j+1)\\
			&\quad<j_U j_D; J M | m_U m_D> <j_U j_D; J' M | m_U m_D>\\
			&\quad<j_1 j; J M^B | m_1^B m^B> <j_1 j; J' M^B | {m_1^B}' m^B>\\
			&\quad\left\{
				\begin{array}{ccc}
					j_1 & j_U-j_1 & j_N \\
					j_D & J & j
				\end{array}
			\right\}
			\left\{
				\begin{array}{ccc}
					j_1 & j_U-j_1 & j_N \\
					j_D & J' & j
				\end{array}
			\right\}\\
			&\quad|j_1 m_1><j_1 m'_1|
	\end{split}
\label{eq:oneBodyDensityMatrixUpB}
\end{equation}

\begin{equation}
	\begin{split}				\rho_{12}(t)&=\sum_{\substack{J,J',M^B,{M^B}'\\j_{12},{j_{12}}',j,m^B\\m_{12}^B,{m_{12}^B}',m_{12},{m_{12}}'}}(-)^{J-J'}\exp\{-it[B(M^B-{M^B}')+g(J(J+1)-J'(J'+1))]\}\\
		&\quad d^J_{M^B,M}(-\beta)d^{J'}_{M',{M^B}'}(-\beta)d^{j_{12}}_{m_{12},m_{12}^B}(-\beta)d^{j_{12}}_{{m_{12}^B}',{m_{12}}'}(\beta')\\		
		&\quad(2j_U+1)(2j_D+1)(2j+1)[(2j_{12}+1)(2j'_{12}+1)]^{\frac{1}{2}}\\
		&\quad<j_U j_D; J M | m_U m_D> <j_U j_D; J' M | m_U m_D>\\
		&\quad<j_{12} j; m_{12}^B m^B |J M^B> <j'_{12} j; {m_{12}^B}' m^B |J' {M^B}'>\\
		&\quad\left\{
			\begin{array}{ccc}
				j_1 & k_U & j_U \\
				j_2 & k_D & j_D\\
				j_{12} & j & J\\
			\end{array}
		\right\}
		\left\{
			\begin{array}{ccc}
				j_1 & k_U & j_U \\
				j_2 & k_D & j_D\\
				j_{12} & j & J'\\
			\end{array}
		\right\}\\
	&\quad|j_{12} m_{12}><j'_{12} m'_{12}|\\
	\end{split}
\label{eq:twoBodyDensityMatrixB}	
\end{equation} 

$\beta$ is the angle of rotation between the flavour axis and $\vec{B}$. Spin projection superscripted with $B$ is the projection in the rotated basis. B is the magnitude of $\vec{B}$. $d^J_{M^B,M}$ is a rotation matrix. All other parameters are as defined in \S{sec:summaryDef}.


\section{Angular Momentum Formulae}

3j-symbol formulae:
\begin{equation}
\label{eq:W3}
W(abcd;ef)\equiv\sum(2c+1)(-)^{f-e-\alpha-\delta}
\left(
			\begin{array}{ccc}
				a & b & e \\
				\alpha & \beta & -\epsilon
			\end{array}
\right)
\left(
			\begin{array}{ccc}
				d & c & e \\
				\delta & \gamma & \epsilon
			\end{array}
\right)
\left(
			\begin{array}{ccc}
				b & d & f \\
				\beta & \delta & -\phi
			\end{array}
\right)
\left(
			\begin{array}{ccc}
				c & a & f \\
				\gamma & \alpha & \phi
			\end{array}
\right)
\end{equation}

\begin{equation}
\label{eq:3j1}
\left(
			\begin{array}{ccc}
				a & b & a+b \\
				\alpha & \delta & \gamma
			\end{array}
\right)=(-)^{a-b-\gamma}\sqrt{\frac{(2a)!(2b)!(a+b+\gamma)!(a+b-\gamma)!}{(2a+2b+1)!(a+\alpha)!(a-\alpha)!(b+\beta)!(b-\beta)!}}
\end{equation}

\begin{equation}
\label{eq:3j2}
\left(
			\begin{array}{ccc}
				a & b & a+b-1 \\
				\alpha & \delta & \gamma
			\end{array}
\right)=(-)^{a-b-\gamma}2(b\alpha-a\beta)\sqrt{\frac{(2a-1)!(2b-1)!(a+b+\gamma-1)!(a+b-\gamma-1)!}{(2a+2b)!(a+\alpha)!(a-\alpha)!(b+\beta)!(b-\beta)!}}
\end{equation}

\begin{equation}
\label{eq:3j3}
\left(
			\begin{array}{ccc}
				a & b & a+b-2 \\
				\alpha & \delta & \gamma
			\end{array}
\right)=(-)^{a-b-\gamma}\sqrt{\frac{(a+b-\gamma-2)!(a+b+\gamma-2)!(2a-2)!(2b-2)!}{2(a-\alpha)!(a+\alpha)!(b+\beta)!(b-\beta)!(2a+2b-1)!}}
\end{equation}

6j-symbol formulae:
\begin{equation}
\label{eq:singleDecoupling}
|(a b)e,d;c>=\sum_{f}|a,(b d)f;c> |(2e+1)(2f+1)]^{\frac{1}{2}}(-)^{a+b+c+d}\left\{
			\begin{array}{ccc}
				a & b & e \\
				d & c & f
			\end{array}\right\}
\end{equation}
\begin{equation}
\label{eq:W6}
\left\{
			\begin{array}{ccc}
				a & b & e \\
				d & c & f
			\end{array}
\right\}=
(-)^{a+b+c+d}W(abcd;ef)
\end{equation}

9j-symbol formulae:
\begin{equation}
\label{eq:doubleDecoupling}
<(ab)c,(de)f; i|(ad)g,(be)h;i> = [(2c+1)(2f+1)(2g+1)(2h+1)]^{\frac{1}{2}}
		\left\{
			\begin{array}{ccc}
				a & b & c \\
				d & e & f\\
				g & h & i\\
			\end{array}
		\right\}
\end{equation}	
\begin{equation}
\label{eq:special09J}
\left\{
\begin{array}{ccc}
	a & b & c \\
	d & e & f\\
	g & h & 0\\
\end{array}
\right\}
=\frac{\delta_{cf}\delta_{gh}(-)^{(c+g-a-e)}}{\sqrt{(2c+1)(2g+1)}}W(abcd;ef)
\end{equation}	
\begin{equation}
\label{eq:special19J}
\left\{
\begin{array}{ccc}
	a & b & c \\
	d & e & c\\
	g & g & 1\\
\end{array}
\right\}
=\frac{a(a+1)-d(d+1)-b(b+1)+e(e+1)}{\sqrt{4c(c+1)(2c+1)g(g+1)(2g+1)}}(-)^{(c+g-a-e)}W(abcd;ef)
\end{equation}
\\
\\
These angular momentum formulae are from \cite{brink68}\cite{thompson94}.


\bibliography{boltzmannFMO}
\bibliographystyle{hplain}

\end{document}